\documentstyle[11pt,newpasp,twoside]{article}
\reversemarginpar
\input psfig

\begin{document}
\title{Intra-Cluster Medium Enrichment and the Dynamical State of
Galaxy Clusters} 
 \author{Sabine Schindler}
\affil{Astrophysics Research Institute, Liverpool John Moores University,
Twelve Quays House, Birkenhead CH41 1LD, U.K.}

\begin{abstract}
The effects of cluster mergers on the metal enrichment of
the intra-cluster gas in clusters of galaxies are reviewed. 
Mergers can influence
the metal production as well as the gas ejection processes,
which transport the gas from the galaxy potential wells into the
intra-cluster gas. Several processes are discussed: ram-pressure
stripping, galactic winds and star formation activity. Simulations
on different scales ranging from galaxy size to large-scale structure
are presented. 
\end{abstract}

\section{Introduction}

X-ray spectra of the intra-cluster gas show metal lines indicating that the gas cannot be only of primordial origin, but a considerable fraction of the gas must have been processed in the cluster galaxies and subsequently been transported from the galaxies into the intra-cluster medium (ICM). 
The amount of metals is
not negligible: the total mass of iron in the ICM is 
of the same order of magnitude as the iron mass in all the galaxies
of a cluster (Mushotzky 1999).
Information about the transport processes can be obtained for example by X-ray observations of clusters. Metal distribution and evolution depend sensitively on the transport processes involved. In some clusters 
the distribution of heavy elements was found to be non-uniform, but 
 increasing towards
the cluster centre (see e.g. De Grandi \& Molendi 2001; De Grandi this
volume). Also ratios of heavy 
elements deviating from solar ratios were found by XMM observations
(e.g. Kaastra et al. 2001). The evolution of heavy elements was hardly measurable with previous X-ray satellites such as ROSAT and ASCA (Schindler 1999), but XMM and CHANDRA finally provide the opportunity to measure metallicities reliably out to redshifts of about $z \approx 1$.

Several processes have been suggested to explain the transport
of gas from the galaxies to the cluster: galactic winds driven by supernovae (De Young 1978),
ram-pressure stripping (Gunn \& Gott 1972), interaction between
individual galaxies and jets from active galaxies. 
Two of these processes, galactic winds  and jets,
are driven by internal processes
within the galaxies. In CHANDRA observations several clusters have
been found in which the pressure of the relativistic particles in the
jets pushes away the ICM leaving depressions in the X-ray emission
(Perseus: Fabian et al. 2000, Hydra-A: McNamara et al. 2000, RBS797:
Schindler et al. 2001). 
The other two processes, ram-pressure stripping  and galaxy-galaxy interaction, are also influenced by their
surroundings, for instance by a cluster merger. Ram-pressure stripping occurs when
a galaxy is moving through the ICM approaching the cluster centre. At
a certain point 
the pressure from the ICM is so strong that the galaxy potential is not deep enough to retain the galactic gas. The gas is stripped off starting
from the outer parts of the galaxy and is lost to the ICM.
Two spectacular examples where the stripping process can be
observed are two galaxies in the Virgo cluster, NGC4501 and
NGC4548 (Cayatte et al. 1990). 

The metal production rate in the galaxies, or  equivalently the star formation activity, is also influenced by external processes. A 
cluster merger, for example, causes a compression in the gas which could trigger a star burst.

\section{Effects of Cluster Mergers}

Mergers of clusters have strong effects both on the physical 
quantities of the cluster and on the observable quantities. 
The most prominent features of mergers are shock waves. The strongest shocks 
emerge after the collision of two clusters. The shocks propagate outwards mainly
in direction of the original collision axis (Schindler \& M\"uller 1993, see
Fig.~1). Also before the actual collision bow shocks can occur.

\begin{figure}[ht]
\psfig{figure=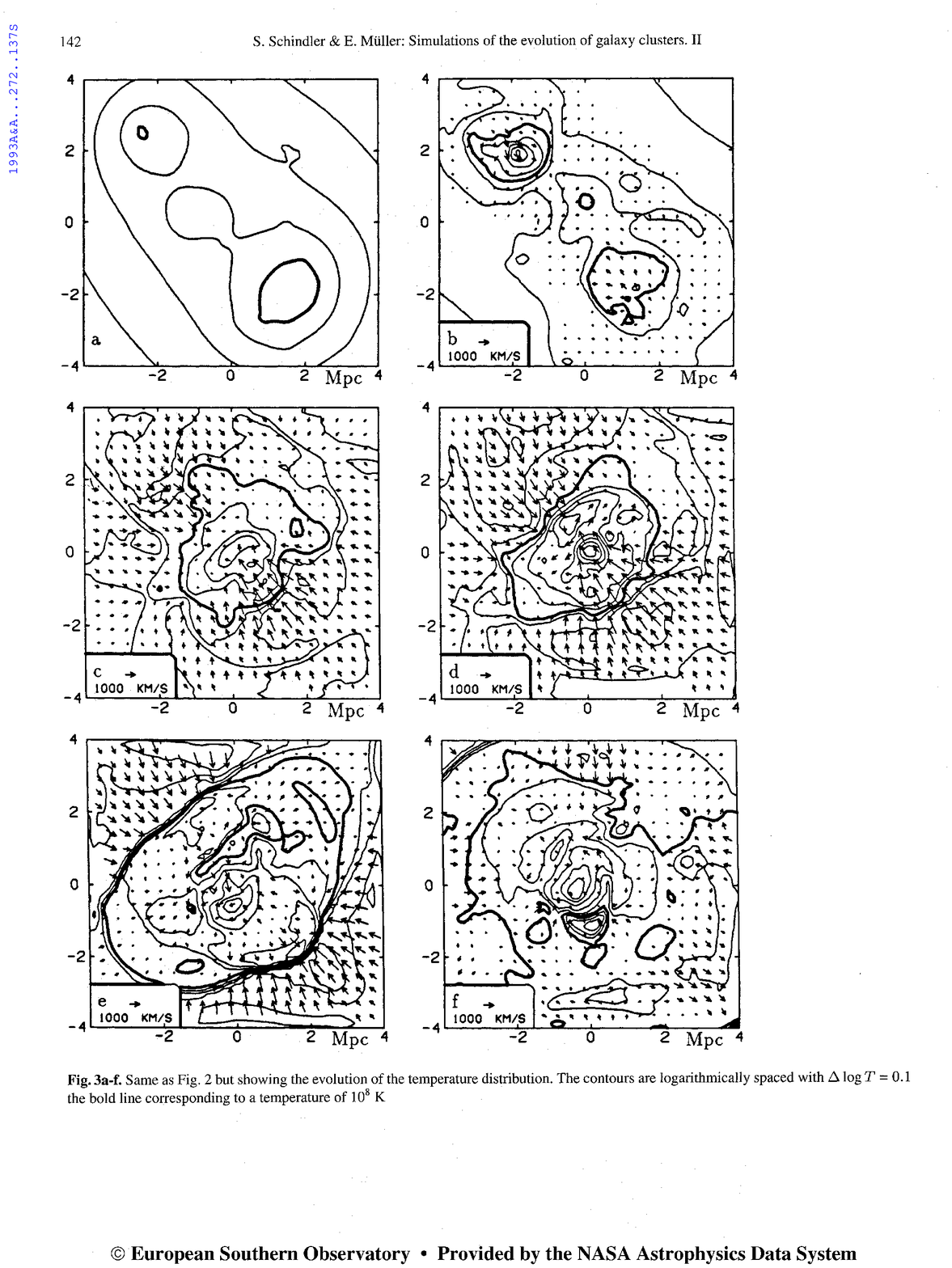,width=10cm,clip=} 
\vskip.2in
\caption{
Temperature contours of the ICM after a major merger. The original
collision axis goes from the upper left corner to the lower right
corner. Two shocks, visible as steep gradients in the temperature, are
propagating outwards along this axis (from Schindler \& M\"uller
1993). 
}
\label{fig:temp}
\end{figure}

\begin{figure}
\psfig{figure=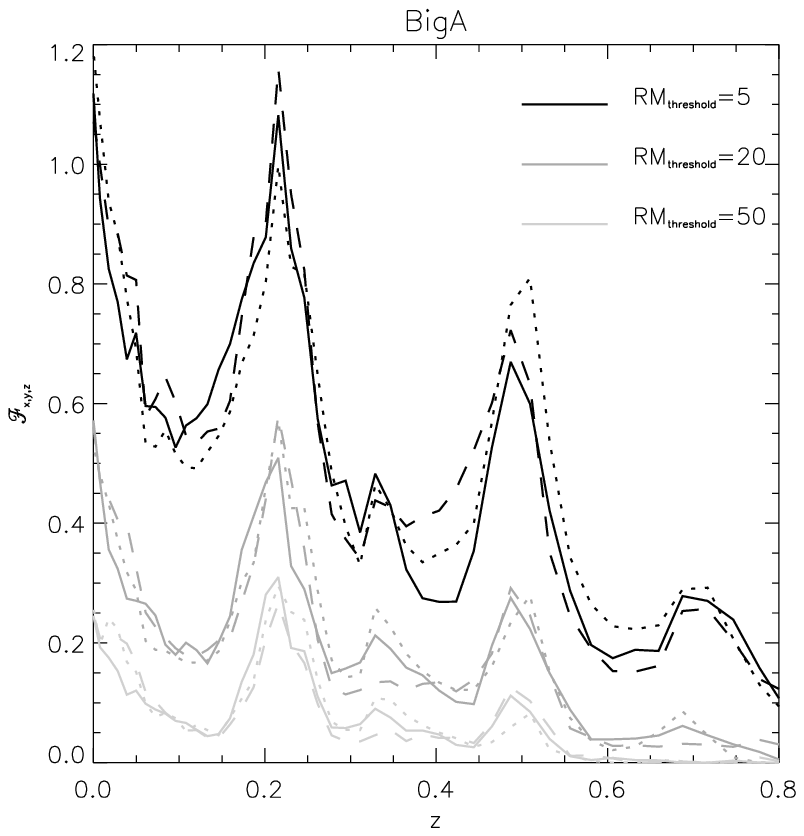,width=12cm,clip=} 
\vskip.2in
\caption{Area exceeding a certain threshold of Faraday 
rotation measure versus redshift $z$ as found in magneto-hydrodynamic
simulations. Different shades of grey correspond to different
thresholds: 5~rad/m$^2$ (black), 20~rad/m$^2$ (dark grey) and
50~rad/m$^2$ (light grey). The area is
normalized to the area emitting 95\% of the cluster X-ray luminosity.
During the core passages of two
subclusters (at $z\approx0.5$ and
$z\approx0.2$) and the subsequent rebounces (at $z\approx0.3$ and at
$z\approx0.0$ and an additional one at  $z\approx0.7$ from an older
core passage) an increase in the rotation
measure, i.e. in the magnetic field, is visible. 
Different projection
directions are shown as different line
types  (from Dolag et al., in
prep.).    
}
\label{fig:klaus}
\end{figure}

The
shocks are not only the major heating source of the inter-cluster gas
and sites of particle acceleration,
but they also change the density and the temperature distribution of
the ICM considerably.
The change in the density distribution is the main influence on the stripping 
rate, because the ram pressure is proportional to $\rho_{ICM} \times
v^2_{rel}$,  
with $\rho_{ICM}$ being the density of the ICM and 
$v_{rel}$ being the relative velocity of the galaxy and the ICM.
Mergers cause also gas motions and turbulence in the ICM. As the ram
pressure depends quadratically on the relative velocity $v_{rel}$ these motions
can increase the stripping rate strongly.

During and after a merger the ICM is mixed by turbulent gas
motions. These motions can change the distribution of heavy elements
within a galaxy cluster. 
Also indirect effects of mergers can influence the
metallicity. Examples of such indirect effects are the increase of the
magnetic field during mergers (see Fig.~2), the increased X-ray luminosity, the 
high number of radio galaxies, and offsets between the ICM
and the dark matter/galaxy distribution.

\section{Influence on the Star Formation Activity}

An important question is whether cluster mergers increase the star
formation rate in galaxies, i.e. the production rate of heavy
elements. 
The star formation rate can be affected by cluster mergers in two
ways.

The interstellar medium in 
a galaxy  can be compressed during a merger, because of the higher
pressure in the surrounding ICM. This compression of the galactic gas would
lead to an increased star formation rate. This effect was predicted 
in simulations by Evrard (1991). Also in a number of observations
a connection between mergers and enhanced star formation rate
has been found, e.g.  in the Coma cluster (Caldwell et al. 1993), 
in A2111 (Wang et al. 1997), 
in A2125 (Owen et al. 1999), and
in several other clusters (Moss \& Whittle 2000).  

In contrast to these results
Fujita et al. (1999) found a decrease of the star formation activity due to a merger. In simulations they see that the interstellar medium in the
galaxies is stripped off  due to the increased ram pressure during 
the merger, so that the galaxies lose most of their gas. Therefore less gas is
left to fuel the star formation process and hence the star formation
activity decreases. Fujita et al. (1999) found an increase of 
post-starburst galaxies at the moment of the subcluster collision,
which indicates that a rapid drop in star formation must have 
occurred.

\section{Simulations of Gas Transport Processes}

Several groups used numerical simulations to investigate the effect of different gas transport processes. In these simulations the
dynamics of the clusters is usually followed by a combination of
N-body and hydrodynamic calculations to account for the different
cluster components: the collisionless components, dark matter and
galaxies, which are best simulated by N-body simulations, and the ICM,
which can be simulated by hydrodynamic methods.

\begin{figure}
\centerline{\psfig{figure=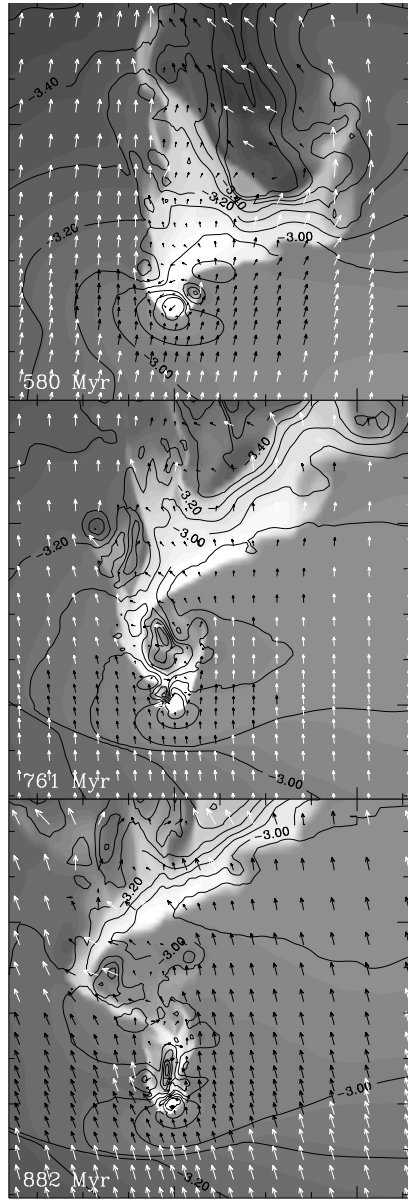,width=5.3cm,clip=}}
\vskip.2in
\caption{Gas density (grey scale) and pressure (contours) of a galaxy
moving downwards towards the cluster centre. The arrows show the Mach
vectors (white when $M>1$, black otherwise). The gas of the galaxy is
stripped due to ram pressure (from Toniazzo \& Schindler 2001). 
}
\label{fig:toniazzo1}
\end{figure}

\subsection{Simulations on Cosmological Scales}

Cosmological simulations accounting for the effects of 
galactic winds were performed e.g. by Cen \& Ostriker (1999). They found 
that the average metallicity increases from 0.01 solar at z=3 to 
0.2  solar at present. The metallicity distribution in their models is not constant but denser regions
have generally higher metal abundances.
Gnedin (1998) took into account not only galactic winds but also 
galaxy-galaxy interactions and concluded that most metals are ejected by
galaxy mergers. In contrast to this result Aguirre et al. (2001) found that 
galaxy-galaxy interactions and ram-pressure stripping are of minor 
importance while galactic winds dominate the metal enrichment of the
ICM. A problem with this kind of simulations is that they must cover a huge range of scales - cosmological scales down to galaxy scales. Therefore the resolution is not
very good at small scales and hence the results have large uncertainties.
This is probably the reason for the discordant results.

\subsection{Simulations with Galactic Winds}

The effects of supernova driven
winds were also investigated with models on cluster 
scales. David  et al. (1991) calculated the first models 
finding that the results depend sensitively in the input parameters:
the stellar initial mass function, the adopted supernova rate and 
the primordial fraction of intra-cluster gas. In the first 
3D models calculating 
the full gas dynamics and the effects of galactic winds on cluster scales
Metzler \&  Evrard (1994, 1997) showed that
winds can account for the observed metal abundances. 
Very strong metallicity gradients were found (almost a factor  of ten between 
cluster centre and virial radius) which are not in agreement with
observations. The authors found that these metallicity gradients 
are hardly affected
by cluster mergers. From simulations on galaxy scales, however, 
Murakami \& Babul  (1999) concluded that galactic winds are 
not very efficient for the metal enrichment in clusters of galaxies.

\subsection{Ram-pressure Stripping on Galaxy Scales}

Another process which is important for the
metal enrichment is ram-pressure stripping: as a galaxy
approaches the cluster centre it experiences an increasingly higher
pressure from the ICM and at some point the galaxy potential is not
strong enough to retain the galactic gas. The gas is stripped
off starting from the outer regions of the galaxies 
and the metals are released into the ICM.

\begin{figure}
\psfig{figure=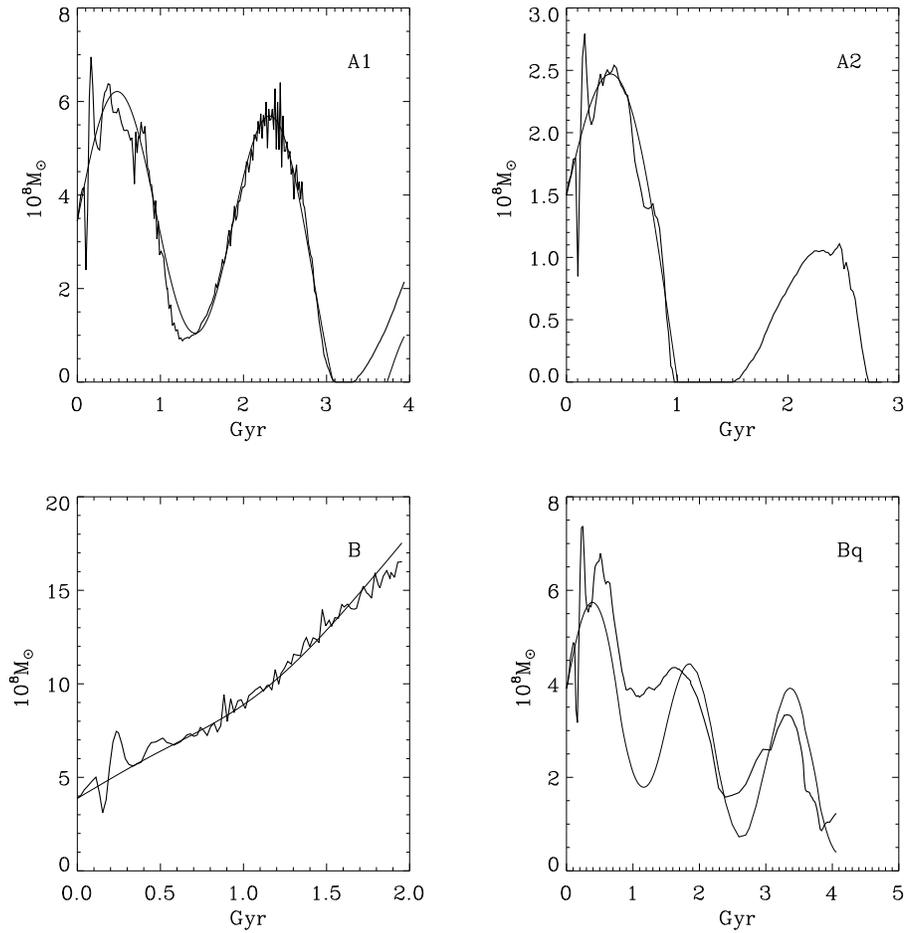,width=13cm,clip=} 
\vskip.2in
\caption{Evolution of the gas mass within the galaxy for different 
galaxy orbits in a cluster. Four different models are displayed which 
show very different behaviour of the gas mass. As the galaxy approaches the cluster centre the gas is 
stripped off  resulting in very low galactic gas masses. In the subsequent apocentric part of the orbit the galaxy can accumulate gas from the surroundings and also from internal processes. Hence the galactic gas mass increases. Therefore a quasiperiodic behaviour can be observed in some 
models (from Toniazzo \& Schindler 2001).}
\label{fig:toniazzo2}
\end{figure}

Simulations of ram-pressure stripping are relatively difficult because 
not only must the conditions of the gas inside the galaxy
and the potential of the galaxy be taken into account, but
also the conditions of the surrounding medium. In several early models
the stripping process was calculated  
(Takeda et al. 1984, Gaetz et al. 1987, Portnoy et al. 1993,
Balsara et al. 1994). 
Recently, high resolution, 3D simulations were
carried out to study the stripping process in different types of
galaxies. Abadi et al. (1999) and 
Quilis et al. (2000) performed simulations of spiral galaxies. They
found that 
the interstellar medium can be stripped off if it is
not homogeneous. For dwarf galaxies 
Mori \& Burkert (2000) found in their simulations (2D) that the gas can be
easily stripped off  when these galaxies move through the intra-cluster
medium. Simulations of elliptical galaxies 
(Fig.~3; 
Toniazzo \& Schindler 2001) showed that also this type of galaxies are
affected by ram-pressure stripping and that the galaxies can even
accumulate  
some gas when they approach 
the apocentres of their orbits (see 
Fig.~4). Also the  X-ray morphologies of
simulated and observed galaxies in the process of stripping are compared.

\begin{figure}
\psfig{figure=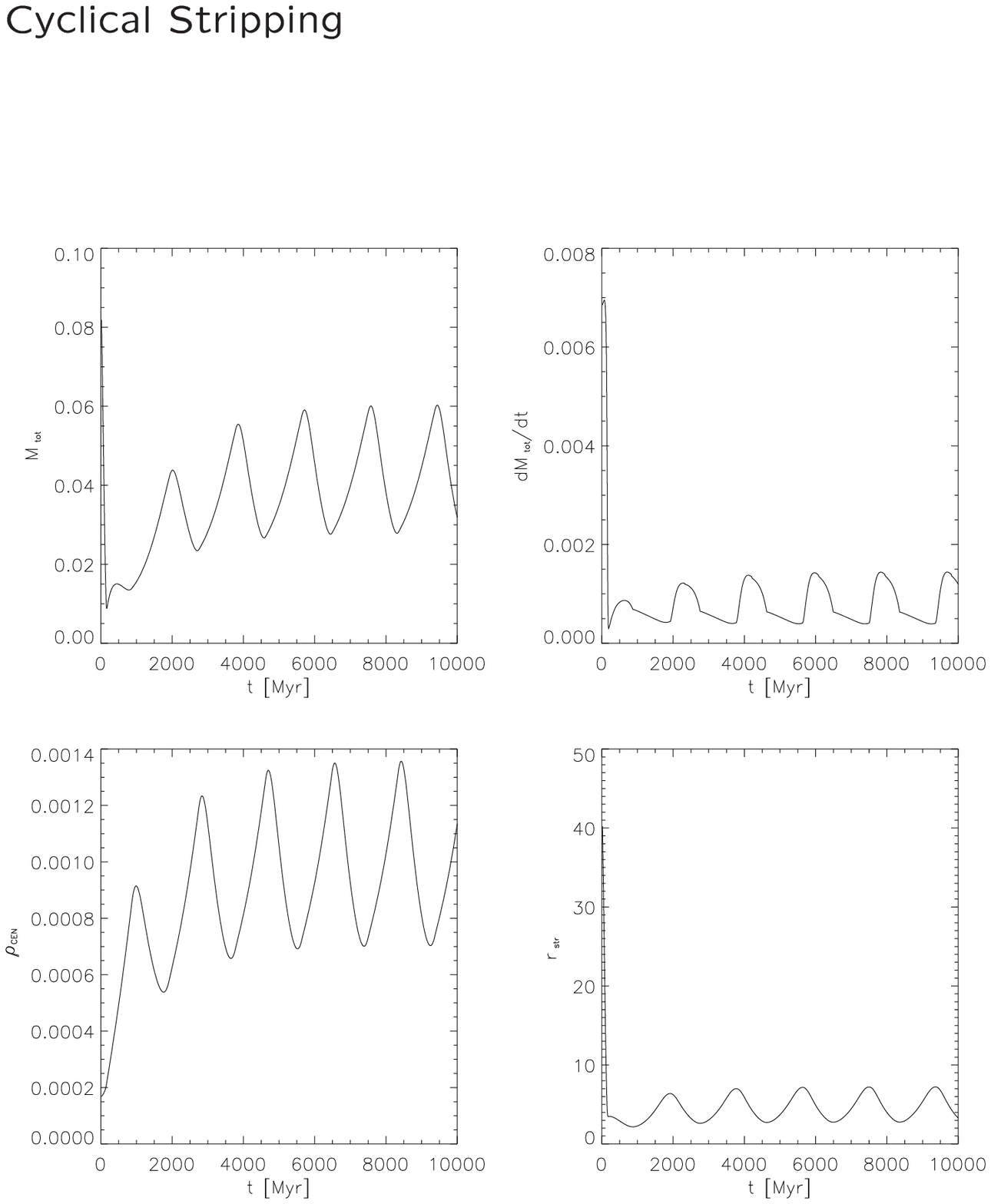,width=13cm,clip=} 
\vskip.2in
\caption{Evolution of the galactic gas mass of an elliptical galaxy in
parametrised form as the galaxy orbits in a cluster 
(from Faltenbacher et al., in prep).}
\label{fig:af}
\end{figure}

All these simulations showed that ram pressure stripping can be
an important metal enrichment process for the ICM. 
Merging activity increases the
effect even more because the ram pressure is proportional to
$\rho_{ICM}\times v_{rel}^2$ with $v_{rel}$ being the relative
velocity of  intra-cluster gas and galaxies. During mergers, not only
is the
ICM density $\rho_{ICM}$ increased but also the relative velocities are higher
than in a relaxed cluster. Therefore a large influence of merging 
processes on the stripping rate is expected. In order to verify and quantify this we are currently performing new comprehensive models on cluster scale which include the stripping process in parametrised form (see Fig.~5). Results will be presented soon.

\section{Summary and Conclusions}

Cluster mergers have important effects on the metal enrichment process
in clusters of galaxies. Both the metal production and the gas
ejection processes are influenced, but so far it is not clear to what
extent. It is still  
controversial whether the star formation rate is increased or
decreased after a merger. It is also controversial which of the gas
transport processes dominate at the different stages of cluster
evolution, how much galactic gas each of the processes can contribute
to the metal enrichment of the ICM and how this contribution is
changed by mergers. Furthermore, it is unclear how much the metal
distribution in the ICM is changed by mergers of subclusters. 
Detailed simulations, including all transport processes and all
cluster components/physics, together with observations of the new
X-ray satellites XMM and CHANDRA, will give final answers soon.

\end{document}